\documentstyle[12pt, rotate, epsf]{article}
\begin{document}
\title{$e^{+}e^{-}\rightarrow (h A)\rightarrow bbbb$ in 
Abelian Extended Supersymmetric Standard Model}
\author{D. A. Demir, N. K. Pak}
\date{{\small Middle East Technical University, Department of Physics, 
06531 Ankara, Turkey}}
\maketitle

\begin{abstract}
We discuss the $e^{+}e^{-}\rightarrow (h A)\rightarrow bbbb$  cross 
section in an Abelian extended SM. We work in that minimum of the scalar 
potential for which Higgs trilier coupling is greater than the soft mass 
parameters. We find that nex-to-lightest Higgs gives the essential 
contribution to the cross section in the small $Z-Z'$ mixing angle and 
leptophobic $Z'$ limit.

\end{abstract}
\newpage
\section{Introduction}

Higgs search is one of the main goals of the present and future 
colliders. specially, in the case o fthe extensions of the SM such as 
MSSM or NMSSM, there are several Higgs particles whose detection in the 
colliders is an important issue. Usually gauge and Yukawa couplings and 
particle masses are unknown, and thus the predictive power of such models 
is limited. Thus, one has to analyze different models to find bounds as 
model independent as possible. 

In this note we shall analyze the indications of Higgs scalars in as 
specific $e^{+}e^{-}$ scattering process in an Abelian extended 
supersymmetric SM. Implications of extra $Z$ bosons appearing in such 
gauge extensions of SM by an extra $U(1)$ have been widely analyzed and 
checked against the precison data, and in the context of the future 
colliders \cite{1,2}.
 
In this work we shall analyze $e^{+}e^{-}\rightarrow (h A)\rightarrow 
bbbb$ cross section in a $U(1)$ extended supersymmetric SM. In 
particular, $e^{+}e^{-}\rightarrow bbbb$ may remind one the recent ALEPH 
four-jet anomaly \cite{3}. However, Higgs interpretation is not 
appropriate to explain the four-jet topology there \cite{3}. For a 
supersymmetrical interpretation of this event one can refer, for example, 
to \cite{4}.

The analysis presented here is based upon the recent work \cite{5}. The 
model under concern is analyzed in detail together with the RGE analysis 
of the parameters of the potential in \cite{5}. There the low energy 
model we discuss is obtained from a supergravity Lagrangian with 
appropriate non-universality in the soft masses at the String scale. Here 
we consider simply a low energy model and summarize some relevant results 
of \cite{5} and derive the necessary quantities for the present problem. 
In particular, we shall work in the trilinear coupling-driven minimum of 
the potential which is discussed in Section 3 and Section 5 of \cite{5}.

This paper is organized as follows. In Section 2 we shall review the main 
results of \cite{5} relevant for this work, and derive the necessary 
quantities for the problem in hand.

In Sec. 3 we shall derive $e^{+}e^{-}\rightarrow (h_{i} A)\rightarrow bbbb$
cross section by using resonance approximation for the scalars. We shall 
base our analysis mainly on the analytical results instead of using 
computer codes such as PHYTIA or JETSET as it was done in \cite{3}.

We will evaluate the cross section and present the variation of the cross 
section against the center of mass energy and Higgs Yukawa coupling. 

Finally in Section 4, a discussion on the results and their implications 
are given.

\section{Higgs Bosons and Vector Bosons}
We will first summarize some of the results of \cite{5}, and derive the 
necessary quantities for the present problem. The gauge group is extended to 
$G=SU(3)_{c}\times SU(2)_{L}\times U(1)_{Y} \times U(1)_{Y'}$ with the 
respective couplings $g_{3}$, $g_{2}$, $g_{Y}$, $g_{Y'}$. We introduce 
the Higgs fields ${\cal{H}}_{1} \sim (1, 2, -1/2, Q_{1})$, ${\cal{H}}_{2} 
\sim (1, 2, 1/2, Q_{2})$, ${\cal{S}}\sim (1, 1, 0, Q_{S})$, with the 
indicated quantum numbers under $G$. The gauge invariance of the 
superpotential 
\begin{eqnarray}
W\ni h_{s}{\cal{S}} {\cal{H}}_{1}\cdot {\cal{H}}_{2}
\end{eqnarray}
guarantees that $Q_{1}+ Q_{2} + Q_{S} = 0$. The scalar potential is given 
by 
\begin{eqnarray}
V(H_{1}, H_{2}, S)&=&m_{1}^{2}|H_{1}|^2+
  m_{2}^{2}|H_{2}|^2+
  m_{S}^{2}|S|^2 - (Ah_{s}SH_{1}\cdot H_{2}+h.c.)\nonumber\\
  &+&|h_{s}|^2\left[ |H_{1}\cdot H_{2}|^2+ |S|^2 ( 
|H_{1}|^2+|H_{2}|^2)\right]\nonumber\\
&+&\frac{G^2}{8}\left(|H_{2}|^2-|H_{1}|^2\right)^2+
   \frac{g_{2}^2}{2}|H_{1}^{\dagger}H_{2}|^2\\&+&
   \frac{g_{Y'}^2}{2}\left(Q_{1}|H_{1}|^2+Q_{2}|H_{2}|^2+
   Q_{S}|S|^2\right)^2\nonumber
\label{pot}
\end{eqnarray}
where $G=\sqrt{g_{2}^{2}+g_{Y}^{2}}$. Here the soft mass parameters 
$m_{1}^{2}$, $m_{2}^{2}$, $m_{S}^{2}$ can have either sign, and without 
loss of generality we choose $h_{S}A$ positive. In terms of the real 
fields $\phi_{i}, \xi_{i}, \psi_{j}\; (i=1,2,3; j=1,2,3,4)$ the Higgs 
fields are defined by 
\begin{eqnarray}
H_{1}&=&\frac{1}{\sqrt{2}}\left(\begin{array}{c c} 
v_{1}+\phi_{1}+i\xi_{1}\\ \psi_{1}+i\psi_{2}\end{array}\right)\\
H_{2}&=&\frac{1}{\sqrt{2}}\left(\begin{array}{c c} \psi_{3}+i\psi_{4}\\ 
v_{2}+\phi_{2}+i\xi_{2}\end{array}\right) \\
S&=&\frac{1}{\sqrt{2}}(v_{s}+\phi_{3}+i\xi_{3})
\end{eqnarray}
Here $\psi_{i}$ determines the charged Higgs sector. 
$v_{1}$, $v_{2}$, $v_{s}$ being real, there is no CP violation at the 
tree level so that the neutral sector of the total scalar mass matrix is 
split into CP-even and CP-odd parts. In that minimum of the potential 
for which  gauge group is completely broken down to color and electric 
symmetries the mass-squared matrices of CP- even and CP-odd scalars 
become, respectively 
\begin{eqnarray}
({\cal{M}}^{2})_{h}=\left(\begin{array}{c c c}
\kappa\frac{v_{s}v_{2}}{v_{1}}+2\lambda_{1}v_{1}^{2}&
-\kappa v_{s}+\lambda_{12}v_{1}v_{2}&-\kappa v_{2} + 
\lambda_{1s}v_{1}v_{s}\vspace{0.1cm}\\ -\kappa v_{s}+\lambda_{12}v_{1}v_{2}&
\kappa\frac{v_{s}v_{1}}{v_{2}}+2\lambda_{2}v_{2}^{2}&
-\kappa v_{1} + \lambda_{2s}v_{2}v_{s}\vspace{0.1cm}\\
-\kappa v_{2} +\lambda_{1s}v_{1}v_{s}&-\kappa v_{1} + \lambda_{2s}v_{2}v_{s}&
\kappa\frac{v_{1}v_{2}}{v_{s}}+2\lambda_{s}v_{s}^{2}\end{array}\right),
\end{eqnarray}
in $(\phi_{1}, \phi_{2}, \phi_{3})$ basis, and 
\begin{eqnarray}
({\cal{M}}^{2})_{A}=\left(\begin{array}{c c c}
\kappa\frac{v_{s}v_{2}}{v_{1}}&
\kappa v_{s}&\kappa v_{2}\vspace{0.1cm}\\ \kappa v_{s}&
\kappa\frac{v_{s}v_{1}}{v_{2}}&\kappa v_{1}\vspace{0.1cm}\\
\kappa v_{2}&\kappa v_{1}&
\kappa\frac{v_{1}v_{2}}{v_{s}}\end{array}\right),
\end{eqnarray}
in $(\xi_{1}, \xi_{2}, \xi_{3})$ basis. Here $\kappa = h_{S}A/\sqrt{2}$, 
$\lambda_{i}=\frac{G^2}{8}+\frac{1}{2}{g_{Y'}}^2Q_{i}^2$, 
$\lambda_{12}=-\frac{G^2}{4}+{g_{Y'}}^2Q_1Q_2+h_s^2$,
$\lambda_{is}={g_{Y'}}^2Q_iQ_S +h_s^2$ and 
$\lambda_{s}=\frac{1}{2}{g_{Y'}}^2Q_S^2$.
The diagonalization of $({\cal{M}}^{2})_{h}$ yields three CP-even scalars
\begin{eqnarray}
h_{i}=(R^{-1})_{ij}\phi_{j}, (i,j =1,2,3).
\label{R}
\end{eqnarray}
The diagonalization of $({\cal{M}}^{2})_{A}$ yields two CP- odd Goldstone 
bosons and a pseudoscalar boson 
\begin{eqnarray}
A^{0}=(F^{-1})_{Aj}\xi_{j}, (j =1,2,3).
\label{F}
\end{eqnarray}

There are two neutral vector bosons: $Z$ boson of $SU(2)_{L}\times U(1)_{Y}$
and $Z'$ boson of $U(1)_{Y'}$ which mix through the mass-squared matrix
\begin{eqnarray}
({\cal{M}}^{2})_{Z-Z'}=\left (\begin{array}{c c}
M_{Z}^{2}&\Delta^{2}\\\Delta^{2}&M_{Z'}^{2}\end{array}\right),
\label{mzz'}
\end{eqnarray}
where
\begin{eqnarray}
M_{Z}^{2}&=&\frac{1}{4}G^2(v_{1}^2+v_{2}^2),\\
M_{Z'}^{2}&=&{g}_{Y'}^{2}(v_{1}^{2}Q_{1}^{2}+v_{2}^{2}Q_{2}^{2}
+v_{s}^{2}Q_{S}^{2}),\\
\label{mix}
\Delta^{2}&=&\frac{1}{2}g_{Y'}\,G(v_{1}^2Q_{1}-v_{2}^2Q_{2}).
\label{delta2}
\end{eqnarray}
\begin{eqnarray}
M^{2}_{Z_{1},Z_{2}}&=&\frac{1}{2}\left[M^{2}_{Z}+M^{2}_{Z'}\mp
   \sqrt{(M^{2}_{Z}-M^{2}_{Z'})^{2}+4\Delta^4}\right],
\end{eqnarray}  
The $Z-Z'$ mixing angle $\alpha$ is given by
\begin{eqnarray}
\alpha=\frac{1}{2}\arctan\left(\frac{2\Delta^2}{M^{2}_{Z'}-M^{2}_{Z}}
\label{mixing}
\right)
\end{eqnarray} 
The coupling of neutral vector bosons to $h_{i}$ and $A^{0}$ can be 
calculated straightforwardly:
\begin{eqnarray}
K^{Z_{1}A^{0}h_{i}}_{\mu}&=&\frac{i}{2}\{(G\cos\alpha- 
2g_{Y'}Q_{1}\sin\alpha)R_{1i}F_{1A}\nonumber\\&-&(G\cos\alpha+
2g_{Y'}Q_{2}\sin\alpha)R_{2i}F_{2A}\nonumber\\&-& 
2g_{Y'}Q_{S}\sin\alpha R_{3i}F_{3A}\}(p_{A}-p_{h_{i}})_{\mu}
\label{zhA1}\\
K^{Z_{2}A^{0}h_{i}}_{\mu}&=&\frac{i}{2}\{(G\sin\alpha+
2g_{Y'}Q_{1}\cos\alpha) R_{1i}F_{1A}\nonumber\\&-&(G\sin\alpha-
2g_{Y'}Q_{2}\cos\alpha) R_{2i}F_{2A}\nonumber\\&+&
2g_{Y'}Q_{S}\cos\alpha R_{3i}F_{3A}\}(p_{A}-p_{h_{i}})_{\mu}
\label{zhA2}
\end{eqnarray}
Moreover, the $b\bar{b}h_{i}$ and $b\bar{b}A^{0}$ vertices are given by 
\begin{eqnarray}
K^{h_{i}bb}&=&\frac{m_{b}}{v_{1}}R_{1i}\\
K^{A^{0}bb}&=&\frac{m_{b}}{v_{1}}i\gamma_{5}F_{1A}
\end{eqnarray}
\section{$e^{+}e^{-}\rightarrow (h_{i} A)\rightarrow bbbb$ cross 
cection} 
We shall calculate the cross section for each possible CP-even neutral 
particle.The scattering process under concern involves four particles in 
the final state. Thus, the phase space integration is too complicated to 
be carried out analytically. We shall calculate the total cross section by 
replacing the $h_{i}$ and $A^{0}$ lines with resonances:
\begin{eqnarray}
\frac{1}{|p_{A}^{2}-mA^{2}+im_{A}\Gamma_{A}|^{2}}&&\rightarrow 
\frac{\pi}{m_{A}\Gamma_{A}}\delta(p_{A}^{2}-m_{A}^{2})\\
\frac{1}{|p_{h_{i}}^{2}-m_{h_{i}}^{2}+im_{h_{i}}\Gamma_{h_{i}}|^{2}} 
&&\rightarrow 
\frac{\pi}{m_{h_{i}}\Gamma_{h_{i}}}\delta(p_{h_{i}}^{2}-m_{h_{i}}^{2})
\label{resonance}
\end{eqnarray}
where $\Gamma_{A}$ and $\Gamma_{h_{i}}$ are the total widths of $A^{0}$ 
and $h_{i}$. Under this approximation one can now calculate the total cross 
section analitycally,
\begin{eqnarray}
\sigma^{i}=\frac{1}{48\pi}s({c^{(i)}_{V}}^{2}+{c^{(i)}_{A}}^{2})\lambda^{3}(1, 
m_{A}^{2}/s, m_{h_{i}}^{2}/s)BR(h_{i}\rightarrow 
b\bar{b})BR(A^{0}\rightarrow b\bar{b}) 
\end{eqnarray}
where 
\begin{eqnarray}
\lambda(x,y,z)=\sqrt{(x-y-z)^{2}-4yz}
\end{eqnarray}
is the phase space factor coming from $Z_{1,2}\rightarrow h_{i}A^{0}$ decay.
The vector and axial couplings are defined by 
\begin{eqnarray}
c^{(i)}_{V}&=&f^{i}_{1}v_{1}+f^{i}_{2}v_{2}\\
c^{(i)}_{A}&=&f^{i}_{1}a_{1}+f^{i}_{2}a_{2}.
\end{eqnarray}
Here the coefficients $f^{i}_{j}$ include the vector boson propagators
and couplings of the vector bosons to scalars
\begin{eqnarray}
f^{i}_{j}=Q^{Z_{j}A^{0}h_{i}}/(s-M^{2}_{Z_{j}})\;\; (i=1,2,3), (j=1,2)
\end{eqnarray}
where, using (16) and (17) we defined $Q^{Z_{j}A^{0}h_{i}}$ via
\begin{eqnarray}
K^{Z_{j}A^{0}h_{i}}_{\mu}=Q^{Z_{j}A^{0}h_{i}}(p_{A}-p_{h_{i}})_{\mu}
\end{eqnarray}
$v_{i}$ and $a_{i}$ in (24) and (25) are given by
\begin{eqnarray}
\left(\begin{array}{c c} v_{1}\\v_{2}\end{array}\right)
&=&\left (\begin{array}{c c}
\cos\alpha&\sin\alpha\\\sin\alpha&-\cos\alpha\end{array}\right)
\left(\begin{array}{c c} v_{e}\\v'_{e}\end{array}\right)\\
\left(\begin{array}{c c} a_{1}\\a_{2}\end{array}\right)
&=&\left (\begin{array}{c c}
\cos\alpha&\sin\alpha\\\sin\alpha&-\cos\alpha\end{array}\right)
\left(\begin{array}{c c} a_{e}\\a'_{e}\end{array}\right)
\end{eqnarray}
where 
\begin{eqnarray}
v_{e}&=&\frac{g_{2}}{4\cos\theta_{W}}(1-4\sin^{2}\theta_{W})\\
a_{e}&=&\frac{g_{2}}{4\cos\theta_{W}}\\
v'_{e}&=&\frac{g_{Y'}}{2}(Q_{L}+Q_{E})\\
a'_{e}&=&\frac{g_{Y'}}{2}(Q_{L}-Q_{E})
\end{eqnarray}
with $Q_{L}$ and $Q_{E}$ being the $U(1)_{Y'}$ charges of lepton doublet 
$L$ and lepton singlet $E^{c}$. 

The minimization of the potential in (\ref{pot}) should be done such that 
the mixing angle (\ref{mixing}) should be below the phenomenological bound 
$\sim 10^{-3}$. Next, if $Z'$ boson is to have an effect on this 
scattering process its mass should be under LEP2 or LHC reach. These two 
constraints can be met under the following conditions:  
\begin{itemize} 
\item If the trilinear coupling $h_{S}A$ is large 
compared to the soft mass parameters then potential
is minimized for
\begin{eqnarray}
v_{1}\sim v_{2}\sim v_{s}\sim 174\,GeV
\label{vevs}
\end{eqnarray}
\item and, if $U(1)_{Y'}$ charges of Higgs doublets satisfy
\begin{eqnarray}
Q_{1}\sim Q_{2}
\end{eqnarray}
so that $\Delta^{2}$, consequently $Z-Z'$ mixing angle becomes small 
without a large $v_{s}$. This yields a vanishingly small $Z-Z'$ 
mixing angle and a relatively light $Z_{2}$. 
\end{itemize} 
In using this procedure we assume that the absolute minimum of the 
potential does not occur in the sfermion sector, which otherwise breaks 
the color and charge symmetries. In this large trilinear coupling limit one 
has definite predictions for the scalar and vector boson masses independent 
of the sign and magnitude of the soft masses in (\ref{pot}): 
\begin{eqnarray}
m_{A}&\sim&\sqrt{\frac{3}{2}}\,h_{S}\,v\\
m_{h_{1}}&\sim&\frac{h_{S}}{\sqrt{2}}\,v\\
m_{h_{2}}&\sim&\frac{1}{2}\sqrt{G^{2}+2h_{S}^{2}}\,v\\
m_{h_{3}}&\sim&\sqrt{3g_{Y'}^{2}Q_{1}^{2}+\frac{h^{2}_{S}}{2}}\,v\\
M_{Z_{1}}&\sim&M_{Z_{0}}\\
M_{Z_{2}}&\sim&\sqrt{3}g_{Y'}Q_{1}\,v\, .
\end{eqnarray}
where $v=246\, GeV$, and the mass spectrum obeys the ordering
$m_{h_{3}}>M_{Z_{2}}>m_{h_{2}}>m_{A}>m_{h_{1}}$, for 
$g_{Y'}^{2}Q_{1}^{2}\sim G^{2}$. 

For any value of the Yukawa coupling $h_{S}$, $m_{A}=\sqrt{3}m_{h_{1}}$. 
This sets an ever- existing gap between the masses of the lightest Higgs 
$h_{1}$ and the pseudoscalar $A^{0}$. 

In the minimum of the potential under concern, the matrices $R$ in 
(\ref{R}) and $F$ in (\ref{F}) become
\begin{eqnarray}
R&=&\left(\begin{array}{c c c} 
\underline{h_{1}}&\underline{h_{2}}&\underline{h_{3}}\vspace{0.1cm}\\
1/\sqrt{3}& -1/\sqrt{2}& -1/\sqrt{6}\vspace{0.1cm}\\
1/\sqrt{3}& 1/\sqrt{2}& -1/\sqrt{6}\vspace{0.1cm}\\
1/\sqrt{3}& 0& \sqrt{2/3}\end{array}\right)
\end{eqnarray}
\begin{eqnarray}
F&=&\left(\begin{array}{c c c}
\underline{A^{2}}&\underline{A^{1}}&\underline{A^{0}}\vspace{0.1cm}\\
-1/\sqrt{2}& -1/\sqrt{6}& 1/\sqrt{3}\vspace{0.1cm}\\
0& \sqrt{2/3}& 1/\sqrt{3}\vspace{0.1cm}\\
1/\sqrt{2}& -\sqrt{2/3}& 1/\sqrt{3}\end{array}\right)
\end{eqnarray}
where $A^{2}$ and $A^{1}$ are the would-be pseudoscalar Goldstone bosons.  

Since $F_{iA}$ and $R_{ih_{1}}$ are identical, one has 
from (\ref{zhA1}) and (\ref{zhA2}),
\begin{eqnarray}
K^{Z_{1}A^{0}h_{1}}_{\mu}&=&0\\
K^{Z_{2}A^{0}h_{1}}_{\mu}&=&0
\end{eqnarray}

Using $R_{ih_{2}}$ and $F_{iA}$ one gets 
\begin{eqnarray}
K^{Z_{1}A^{0}h_{2}}_{\mu}&=&-\frac{1}{\sqrt{6}}G\cos\alpha 
(p_{A}-p_{h_{3}})_{\mu}\\ K^{Z_{2}A^{0}h_{2}}_{\mu}&=&0
\end{eqnarray} 
where in demonstrating that $K^{Z_{2}A^{0}h_{2}}_{\mu}$ vanishes, 
we used the equality of the $U(1)_{Y'}$ charges of the doublets.

Finally, using $R_{ih_{3}}$ and $F_{iA}$ one gets
\begin{eqnarray}
K^{Z_{1}A^{0}h_{3}}_{\mu}&=&-\frac{1}{\sqrt{2}}g_{Y'}Q_{S}\sin\alpha 
(p_{A}-p_{h_{3}})_{\mu}\\
K^{Z_{2}A^{0}h_{3}}_{\mu}&=&\frac{1}{\sqrt{2}}g_{Y'}Q_{S}\cos\alpha
(p_{A}-p_{h_{3}})_{\mu}
\end{eqnarray}
which reflect completely the extended nature of the model.

Now we shall discuss the implications of the vector boson - Higgs 
couplings in (44)-(49). As is seen from (44) and (45) there is no 
coupling of vector bosons to the lightest Higgs scalar and the 
pseudoscalar boson. These two equations guarantee that the $U(1)_{Y'}$ 
extended model under concern behaves very much like the SM in the large 
trilinear coupling driven minimum. In particular, (44) is a reminder of 
the SM where there is no pseudoscalar boson at all. As dictated by (45), 
in this minimum of the potential $Z_{2}$ is also similar to $Z$ in this 
respect. This vanishing of the coupling constants is important in that 
$h_{1}$ does not contribute to the total cross section. 

The coupling of the vector bosons to next-to-lightest Higgs $h_{2}$ and 
$A^{0}$ are given by (46) and (47). As is seen from (46) $Z_{1}$ now 
feels $h_{2}$ and $A^{0}$ by a non-zero coupling constant 
$-\frac{1}{\sqrt{6}}G\cos\alpha$. As we require the mixing angle be small 
($\cos\alpha \sim 1$) this coupling constant is no way negligable in the 
minimum of the potential under concern. The nature of the coupling is 
essentially weak, since extended nature of the model enters ony by the 
$Z-Z'$ mixing angle.

The coupling of the vector bosons to heaviest Higgs $h_{3}$ and pseudoscalar
$A^{0}$ are given by (48) and (49). In this case both $Z_{1}$ and $Z_{2}$ 
couple to scalars. It is for $h_{3}$ case that the couplings carry the 
seeds of the extended nature of the model. Strengths of the couplings are
proportional to the $U(1)_{Y'}$ coupling constant times the $U(1)_{Y'}$ 
charge of the singlet $S$. Unless $g_{Y'}Q_{S}$ is unnaturally large as 
compared to the moderate choice $g_{Y'}Q_{S}\sim G$, in the small $Z-Z'$ 
mixing angle limit $Z_{1}$ essentially decouples from scalars leaving 
room only for $Z_{2}$. 

A number of authors have explained a possible excess in $R_{b}$ by a 
leptophobic $Z'$ \cite{6}. The most recent LEP data \cite{7} weakened the 
possibility of an  $R_{b}$ excess. So $Z'$, if exists, would probably be 
hadrophobic. Thus we are to take $Z'$ as leptophobic as possibile as 
required by the present data. 

We can summarize the situation concerning the process under consideration 
(under the resonance approximation) as follows. $h_{1}$ is excluded from 
the process by (44) and (45) and, consequently it brings no constraint on 
Yukawa and  gauge couplings. As we see from (46)-(49), $Z_{2}$ 
does not contribute to the scattering process until $s$ exceeds the 
kinematical threshold $(m_{A}+m_{h_{3}})^{2}>  (m_{A}+M_{Z_{2}})^{2}$. 
$m_{h_{3}}$ is bounded by $M_{Z_{2}}$ from below, and the latter depends 
on the $U(1)_{Y'}$ coupling constant and  $U(1)_{Y'}$ charge of the 
singlet $S$. If $M_{Z_{2}}$ is beyond the LEP2 reach so does $m_{h_{3}}$.
  
The axial and vector couplings of the vector bosons are given in (24) and 
(25). From these equations, and from (46) and (47), it follows that
\begin{eqnarray}
c_{V}^{(2)}&=&-\frac{1}{\sqrt{6}}G\cos\alpha 
\frac{1}{s-M_{Z_{1}}^{2}}(\cos\alpha v_{e} +\sin\alpha v'_{e})\\
c_{A}^{(2)}&=&-\frac{1}{\sqrt{6}}G\cos\alpha 
\frac{1}{s-M_{Z_{1}}^{2}}(\cos\alpha a_{e} +\sin\alpha a'_{e})
\end{eqnarray}
In the case of small $Z-Z'$ mixing angle, which is really the case in the 
minimum of the potential, $U(1)_{Y'}$ contribution to 
vector and axial couplings of leptons is suppressed by $\sin\alpha$. 
Therefore, the vector and axial couplings in (50) and (51) practically do 
not get any significant contribution from leptonic $U(1)_{Y'}$ charges; 
\begin{eqnarray}
c_{V}^{(2)}&\sim& -\frac{G}{\sqrt{6}}\frac{v_{e}}{s-M_{Z_{1}}^{2}}\\ 
c_{A}^{(2)}&\sim& -\frac{G}{\sqrt{6}}\frac{a_{e}}{s-M_{Z_{1}}^{2}}. 
\end{eqnarray}
For moderate values of $g_{Y'}Q_{S}$,  ($g_{Y'}Q_{S}\sim G$), in the small 
$Z-Z'$ mixing angle limit, one can neglect coupling of $Z_{1}$ to 
$h_{3}$ and $A^{0}$ in (48). Then $Z_{2}$ couples to scalars by a non 
negligable coupling constant $\sim \frac{1}{\sqrt{2}}g_{Y'}Q_{S}$, as is
seen from (49). Under this approximation, the leptonic couplings are 
given by
\begin{eqnarray}
c_{V}^{(3)}&\sim&-\frac{1}{\sqrt{2}}g_{Y'}Q_{S}
\frac{1}{s-M_{Z_{2}}^{2}}v'_{e}\\
c_{V}^{(3)}&\sim&-\frac{1}{\sqrt{2}}g_{Y'}Q_{S}
\frac{1}{s-M_{Z_{2}}^{2}}a'_{e}
\end{eqnarray}
which can be large enough to make $Z_{2}$ effects be seen in the 
present-day experiments. Altough it was not the case in the $h_{1}$ and 
$h_{2}$ couplings, here one has to choose $Z'$ as leptophobic as 
required by the experiment. From the form of the $v'_{e}$ and $a'_{e}$ 
given in (32) and (33), we conclude that $U(1)_{Y'}$ charges of both $L$ 
and $E^{c}$ must be choosen small. This makes $Z_{2}$ to be hardly 
observable. 

Depending on the leptophobicity of $Z'$ to suppress the vector and axial 
couplings in (54) and (55), one concludes that only $h_{2}$ gives a 
significant contribution to the total cross section through (46). This is 
an important result in that one can single out the next-to-lightest Higgs 
among others by measuring the cross section. Reading $m_{A}$ and 
$m_{h_{2}}$ from (36) and (38), and using (52) and (53) we can rewrite the 
total cross section as follows 
\begin{eqnarray}
\sigma_{tot}\simeq \sigma^{(2)}= 
\frac{1}{72\pi}\frac{1}{v^{2}}(v_{e}^{2}+a_{e}^{2})\frac{r_{Z}f(r_{Z}, 
r_{S})}{(1-r_{Z})^{2}} BR(h_{2}\rightarrow 
\bar{b}b)BR(A^{0}\rightarrow \bar{b}b) 
\end{eqnarray}
where we introduced the definitions
\begin{eqnarray}
r_{Z}&=&M_{Z_{1}}^{2}/s\simeq M_{Z_{0}}^{2}/s\\
r_{S}&=&h_{S}^{2}/G^{2}\equiv \mu^{2}/M_{Z_{0}}^{2}\\
f(r_{Z},r_{S})&=&\{1+(4r_{S}-1)^{2}r_{Z}^{2}-2(8r_{S}+1)r_{Z}\}^{3/2} 
\end{eqnarray}
and effective $\mu$ parameter is defined by $\mu= h_{S}v_{s}/\sqrt{2}$.
Thus, cross section depends on two variables only: $r_{Z}$ and $r_{S}$, in 
particular, it does not depend upon the $U(1)_{Y'}$ coupling constant and 
$U(1)_{Y'}$ charges of leptons and Higgs fields. 

In Fig. 1 we present the dependence of the cross section $\sigma$ on the 
$r_S$ and $r_Z$ for particular values of the branching ratios 
$BR(h_{2}\rightarrow b\bar{b})\sim 0.8$ and $BR(A^{0}\rightarrow 
b\bar{b})\sim 0.8$. When the scalars are heavy enough the dominant decay 
mode is $\bar{b}b$ (including the gluonic final states). Under the 
present lower bounds on the scalar masses (which are mostly 
model dependent) $b\bar{b}$ dominance is guaranteed so that small $r_{S}$ 
portion of Fig. 1 is irrelevant.

In Fig. 1, $r_Z$ is allowed to vary from 0.15 
$(\sqrt{s}\sim 240\, GeV)$ to 0.5 $(\sqrt{s}\sim 130\, GeV)$. At the 
latter end, cross section is bigger due to the closeness of this end to 
the $Z_{0}$ pole. As $s$ grows to larger values the cross section falls 
gradually and becomes numerically $\sim 0.2 pb$ around $(\sqrt{s}\sim 
240\, GeV)$. When $s$ is below the kinematical threshold of 
$(m_{A}+m_{h_{2}})^{2}$, $Z_{i}\rightarrow A^{0}h_{2}$ is forbidden, and 
we plot in these regions $\sigma =0$ surface to form a reference level. 
In an exact treatment of the process, this region would be smaller 
since the resonance approximation puts the restriction of reality on 
the scalars whereby narrowing the available phase space. As wee observe 
from the figure for higher values of $s$ the 
region of 'non-zero' cross section becomes wider, making observability 
possible. Actually, the phenomenological bounds on the pseudocalar mass 
must have been taken into account, however as these bounds might change, we 
present the plot in entire $h_{S}$ range to show the variation of cross 
section with $h_{S}$ and $s$.

Let us note that, in the above discussion, we have concentrated on the 
coupling of the  vector bosons to the CP-even and CP-odd scalars to 
analyze the scattering process under concern. Altough this discussion is 
sufficient for the aim of the work, one could analyze other modes of 
Higgs production, such as the Higgs strahlung process \cite{8}, 
$e^{+}e^{-}\rightarrow h_{j}Z_{k}$ in the model under consideration. To 
give an idea, we shall list the $Z_{i}Z_{j}h_{k}$ coupling when 
$g_{Y'}Q_{S}\sim G$ and mixing angle is negligable small
\begin{eqnarray}
K_{h_{1}Z_{1\mu}Z_{1\nu}}&=&\frac{v}{\sqrt{24}}G^{2}g_{\mu\nu},\;\; 
K_{h_{3}Z_{1\mu}Z_{1\nu}}=\frac{v}{\sqrt{48}}G^{2}g_{\mu\nu},\;\;
K_{h_{1}Z_{2\mu}Z_{2\nu}}=\frac{v}{\sqrt{24}} 3 
g_{Y'}^{2}Q_{S}^{2}g_{\mu\nu},\nonumber\\
K_{h_{3}Z_{2\mu}Z_{2\nu}}&=&\frac{v}{\sqrt{48}} 3 
g_{Y'}^{2}Q_{S}^{2}g_{\mu\nu},\;\;
K_{h_{2}Z_{1\mu}Z_{2\nu}}=\frac{v}{2}Gg_{Y'}Q_{S}^{2}g_{\mu\nu}
\end{eqnarray}
The remaining couplings are proportional to $\sin\alpha$, and thus small 
compared to those in (60), $g_{Y'}Q_{S}\sim G$. As a result, coupling of 
$Z_{i}Z_{j}$ to $h_{2}$ is negligably small compared to others. There is 
again a profound difference between $h_{2}$ and others. Namely, unlike 
$h_{1,3}$, $h_{2}$ shifts $Z_{1}$ to $Z_{2}$ and vice versa, which allows 
one to single out $h_{2}$ among others. Higgs search through 
Higgs-strahlung channel will be discussed elsewhere \cite{9}. 

\section{Conclusions and Discussions} 
 
The smallness of the mixing angle is an essential phenomenological 
restriction on such gauge extensions of MSSM \cite{1,2}. Furthermore, 
leptophobicity 
is an indispensable requirement according to LEP results. Under these two 
requirements, we have analyzed the $e^{+}e^{-}\rightarrow (h_{i} 
A)\rightarrow bbbb$ scattering in an Abelian extended SM.

In the large triliear coupling limit, model yields interesting results in 
that only next-to-lightest Higgs contribute to the process. The $Z'$ 
boson is essentially unobservable as far as the leptonic current is 
considered. In this sense model is similar to NMSSM \cite{10} where one 
extends the 
Higgs sector with a singlet. However, determination of the parameters 
requires a different analysis which is outside the scope of this work.

We have analyzed the present model at the tree-level. At the loop level, 
behaviour of the potential, and thus, all the physical parameters derived 
here,  including the mixing angle itself, would naturally change \cite{11}.
Strength of the variations in the potential parameters depends on the 
$U(1)_{Y'}$ charges of the particle spectrum, Yukawa couplings 
(especially the top Yukawa coupling), gauge couplings and soft 
parameters. Here we limited our work to tree level analysis, leaving the 
consideration of the radiative corrections to another work. 

In near future, LEP II, LHC or NLC may catch the signals of $Z'$ boson 
after which the predictions or the assumptions of this work can be tested 
against the experimental results. For example, after a two year run, LEP II
will be able to count up to $\sim 150$ events depending on the lwer bound 
on the scalar masses. 

One of us (D. A. D. ) thanks T. M. Aliev for discussions.

\newpage
\begin{figure}
\vspace{12.0cm}
    \includegraphics{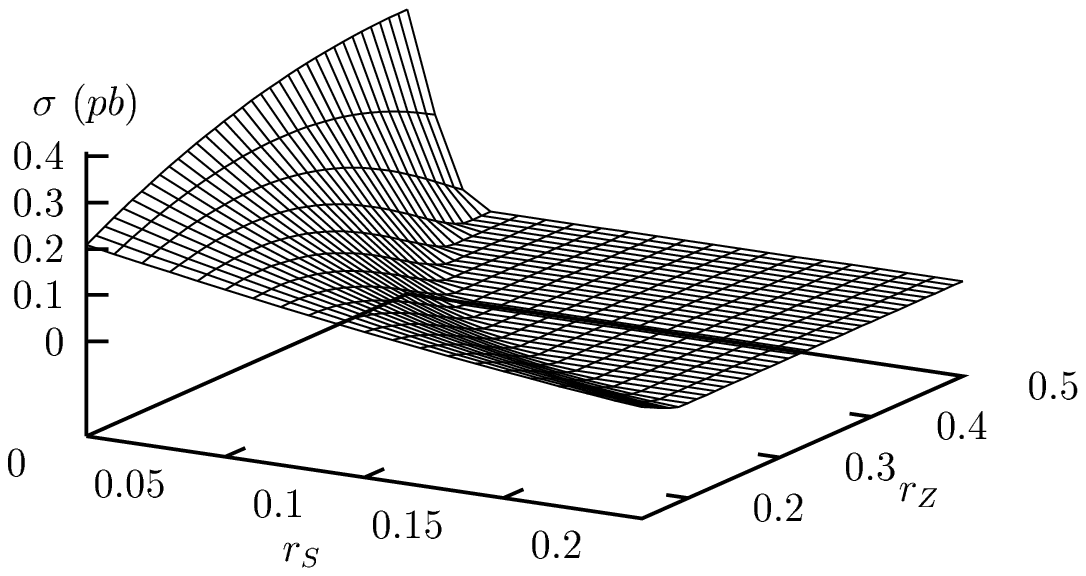}
    \vspace{-8.0cm}
\vspace{0.0cm}
\mbox{ \hspace{0.5cm} \large{\bf Figure 1: Dependence of $\sigma$ on
$r_{S}$ and $r_{Z}$}}
\end{figure}

\begin{thebibliography}{99}
\bibitem{1} {M. Cvetic, P. Langacker, Mod. Phys. Lett. A 11 (1996) 1247.}
\bibitem{2} {M. Cvetic, S. Godfrey, hep-ph/9504216.}
\bibitem{3} {ALEPH Collab. D. Buskulic et. al., Z. Phys. C 71 (1996) 197.}
\bibitem{4} {M. Carena, G. F. Guidice, S. Lola, C. E. M. Wagner, Phys. 
Lett. B 395 (1996) 225.}
\bibitem{5} {M. Cvetic, D. A. Demir, J. R. Espinosa, L. Everett, P. 
Langacker, hep-ph/9703317.}
\bibitem{6} {See, for example, V. Barger, K. Cheung, P. Langacker, Phys. 
Lett. B 381 (1996) 226.}
\bibitem{7} {LEP Electroweak Working Group and SLD Heavy Flavour Group, 
CERN-PPE/96-183.}
\bibitem{8} {See, for example, F. de Campos et. al. Phys. Lett. B 336 
(1994) 446.}
\bibitem{9} {D. A. Demir, N. K. Pak (work in progress)} 
\bibitem{10} {See, for example, S. F. King, P. L. White, Phys. Rev. D 52 
(1995) 4183.}
\bibitem{11} {See, for example, M. Quiros, hep-ph/9703412.}
 \end{thebibliography}
\end{document}